\documentclass[[aps,prx,twocolumn,superscriptaddress,floatfix,amsmath,amssymb]{revtex4} 
\usepackage{graphicx} 
\usepackage{Jonasmacros} 
\usepackage{bm}
\usepackage{mathptmx}
\usepackage{txfonts}
\usepackage{epstopdf}

\usepackage[colorlinks,citecolor=blue,linkcolor=magenta]{hyperref}
\usepackage[usenames,dvipsnames,svgnames]{xcolor}

\begin{document}
\title{ Time-dependent potential impurity in topological insulator }
\author{Saurabh Pradhan}
\affiliation{Department of Physics and Astronomy, Box 516, 75120, Uppsala University, Uppsala, Sweden}
\affiliation{Institut f\"{u}r Experimentalphysik, Freie Universit\"{a}t Berlin,
Arnimallee 14, 14195 Berlin, Germany}
\author{Jonas Fransson}
\affiliation{Department of Physics and Astronomy, Box 516, 75120, Uppsala University, Uppsala, Sweden}

\begin{abstract}
We consider periodically driven potential impurities coupled to the surface states of a two-dimensional topological insulator. The problem is addressed by means of two models, out which the first model is an effective continuum Hamiltonian for the surface states, whereas the Kane-Mele lattice model is our second approach. While both models result in drastic changes in the local density of electron states with increasing amplitude and frequency of the driving field, the linearly low energy local density of electron states remains in the continuum model, however, with an increased Fermi velocity. The spectrum of the continuum model remains gapless under the emergence of new impurity resonances near the Fermi energy. The Kane-Mele lattice model represents a finite size system, with edge states appearing at the boundary of the system. We, thus, consider the impurity at two different positions, one at the boundary and one at the center of the lattice. In the former case, a reduction and broadening of the low energy local density of electron states result with increasing amplitude of the driving field. On the other hand, there are no new resonances emerging in the spectrum. In the latter case, the spectrum is gapped both in the absence of the impurity as well as for weak amplitudes of the driving field, while the gap tends to fill up with impurity states with increasing amplitude.
\end{abstract}
\maketitle

\section{Introduction}
Classical and quantum systems show new phases with unexpected properties when driven by an external time-dependent periodic field. 
In many cases, these phases do not have any counterpart in static systems. 
For example, periodic driving in a one dimensional
systems generates chaotic motion of particles \cite{chaos1,chaos2,chaos3,chaos4}.
In molecular systems, temporal changes of microscopic parameters or bias voltage
may lead to various exotic non-equilibrium states \cite{Fainshtein}. The ratio between kinetic and potential energy as well as lattice spacing can be
varied in optical lattices to create non-trivial states \cite{coldatom1, 
coldatom2}. 
Recently, periodically driven systems have
given rise new phases \cite{ drive1,drive2,drive3,drive4,drive5,drive6,drive7,drive8,
drive9,drive10}, such
as, Floquet block-states in topological insulator \cite{ftimat1,ftimat2},
optical lattices \cite{ftiopt}, and cold-atom \cite{fticold} systems.
Topological insulators have metallic edge states due to the spin-orbit coupling
and non-trivial topology of the band structure, while sustaining an insulating gap in the bulk. Material such as
Bi\textsubscript{2}Se\textsubscript{3} and Bi\textsubscript{2}Si\textsubscript{3} shows topologically protected Dirac cone, resulting from the properties of these edge states \cite{ti1,ti2,ti3}. Moreover, due to spin-momentum locking, back-scattering off potential impurities is prevented. In other words, the definite chirality of the edge states does not allow transitions between the states $|{\bf k}\uparrow\rangle$ and $|-{\bf k}\downarrow\rangle$ assisted by scattering off time-reversal symmetric (non-magnetic) impurities.

The robustness of these edge mode has been proven theoretically by examining the 
backscattering due to potential impurity in continuum and lattice model 
\cite{edge1, edge2, edge3,edge4,edge5,edge6}. The protection of the Dirac 
surface states against disorder is also seen in various experiments 
\cite{edge7,edge8,edge9}. Magnetic impurities, on the other hand, do not keep 
these edge modes intact which results in imperfect quantization of the 
conductance \cite{mag-im1, mag-im2, mag-im3}. 
 
Even though there are many theoretical and experimental 
studies of the effects of static  impurities on topological insulator, the effects of dynamical impurities have not been considered previously. 
Dynamical local interaction can give rise to various non-linear phenomena such 
as multi-photon dissociation or excitation of
atom or molecule when exposed under strong laser field \cite{lesser1, 
lesser2,lesser3}.
In this article, we consider an atomically sharp time-dependent impurity potential, simulating the conditions of a local, yet extremely focused, monochromatic laser field, affecting a single lattice site. We show the emergence of impurity resonances, in accordance with previously discussed for impurity resonances for stationary conditions \cite{edge4,edge5,edge6,PhysRevB.94.075401}, which give rise to strong modifications of the local density of electron states near the impurity site, with an increasing number of resonance features with increasing amplitude and frequency of the driving field. While the metallic edge states tend to develop a sharp density peak around the Fermi level for impurities interacting with the edge states, the density gap of the bulk states become increasingly filled up with increasing scattering potential strength, for impurities directly perturbing the bulk sites.

\section{Model and Floquet-Green Function}
In this section, we describe the general formalism of the Floquet Green function 
method. The Hamiltonian of the full quantum system is a periodic function in 
time, $H(t) = H(t+\tau)$, where $\tau  = 2\pi/\Omega$ is the period of the 
external driving field. As a direct consequence of the periodicity, we can use 
the Floquet theorem \cite{Floquet1,Floquet2,Floquet3} which can be regarded as 
a time domain equivalent to the Bloch theorem.
Due to the explicit time-dependence of the Hamiltonian, electrons can be excited to different energy states. However, if the time-dependent potential is characterized by a single frequency ($\Omega$), the energy difference between the final and initial state should be an integer multiple of $\Omega$. This restriction gives rise to an energy space representation of the Hamiltonian, the corresponding Green function, as well as of operators.

We define the Floquet Hamiltonian, a Hermitian operator, for the generic time-dependent Hamiltonian
$H(x,t)$
\begin{align}
 H^F(x,t) = H(x,t) - i\hbar\frac{\partial}{\partial t}
 	,
\end{align}
which gives the Floquet green function written as
\begin{align}
\Bigl(
	\epsilon - H^F(x,t)
	\Bigr)
	G^F(x,x';t,t')=\delta(x-x')\delta(t-t')
	.
\end{align}
The time periodicity of the Hamiltonian is inherited in the Floquet Green
function, which becomes periodic in both $t$ and $t'$. Hence, we can Fourier expand both the Green function and Hamiltonian into
\begin{subequations}
\begin{align}
H^F(x,t)=&
	\sum_\gamma
		H^F_{\gamma}(x)e^{i\epsilon_\gamma t}
	,
\\
G^F(x,x';t,t')=&
	\sum_{\alpha,\beta}
		G_{\alpha\beta}^F(x,x')e^{i\epsilon_\alpha t-i\epsilon_\beta t'}
	.
\end{align}
\end{subequations}
Here, the quasi-energy $\epsilon_\alpha$ is a conserved quantity. In this way, a time-periodic driven system is reduced to an algebraic matrix equation. 
Although the Fourier expansion reduces the complexity of the problem, the dimensions of the corresponding Hilbert space is infinite. Therefore, there is no exact analytical closed form of the Green function. In fact, even the exact solution of a two-level system driven with linearly polarized light is not known \cite{shirley}.

Due to this intrinsic complexity, we approach the Floquet Green function numerically, by considering a harmonic monochromatic time-dependent driving field. The Hamiltonian, then, assumes the form
\begin{align}
 \mathbf{H} = \mathbf{H}^0 + 2\mathbf{V}\cos\Omega t
 	,
\end{align}
where $\mathbf{H}^0$ is the time-independent part of the Hamiltonian whereas $\mathbf{V}$ represents the coupling to the time-dependent driving field. The corresponding Floquet Hamiltonian for the assumed driving field has block tri-diagonal structure, according to
\begin{align}
H^F_{\alpha\beta}=
	(\mathbf{H}^0 - \alpha\hbar\Omega)\delta_{\alpha\beta}
	+
	\mathbf{V}(\delta_{\alpha+1\beta} + \delta_{\alpha-1\beta})
\end{align}
The Floquet green function becomes
\begin{align}
(\mathbf{1}E_\alpha - \mathbf{H}^0)G_{\alpha\beta}
	-
	\mathbf{V}(G_{\alpha+1\beta}
	+
	G_{\alpha-1\beta})
	=
	\mathbf{1}\delta_{\alpha\beta}
	,
\end{align}
where $E_\alpha = \epsilon +\alpha\hbar\Omega$. This matrix equation can be 
solved iteratively by using the \emph{matrix continued fraction} method 
\cite{martinez1,martinez2}. As a result, we obtain a recursive matrix equation 
for the Green function
\begin{subequations}
\begin{align}
\Big[\mathbf{1}E_\alpha
	-
	\mathbf{H}^0
	-
	\mathbf{V}_\text{eff}(E_{\alpha})\Big]G_{\alpha,\alpha}
	=&
	\mathbf{1}
	,
\\
\mathbf{V}_\text{eff}(E_{\alpha})=&
	 \mathbf{V}^+_\text{eff}(E_{\alpha})
	 +
	 \mathbf{V}^-_\text{eff}(E_{\alpha})
	 .
\end{align}
\end{subequations}
Here, the effective potential is given by
\begin{align}
\mathbf{V}^\pm_\text{eff}(E_{\alpha})=&
	\mathbf{V}\frac{1}{\mathbf{C}_{\alpha\pm1}
	-
	\mathbf{V}\frac{1}{\mathbf{C}_{\alpha\pm2}
	-
	\mathbf{V}\frac{1}{\vdots}\mathbf{V}}\mathbf{V}}\mathbf{V}
	,
\end{align}
where $\mathbf{C}_{\alpha} = \mathbf{1}E_{\alpha} - \mathbf{H}^0$.
Assuming that the impurity sit at the origin $(\mathbf{r} = 0)$, the effective potential can be written
\begin{align}
\mathbf{V}^\pm_\text{eff}(E_{\alpha})=&
	|0\rangle\langle 0|
		\frac{1}{\mathbf{C}_{\alpha\pm1}-|0\rangle\langle 0|\frac{1}{\mathbf{C}_{\alpha\pm2}-\frac{1}{\vdots}}|0\rangle\langle 0|}
	|0\rangle\langle 0|
\nonumber\\=&
	|0\rangle\langle 0|
		\frac{1}{\mathbf{C}_{\alpha\pm1}-|0\rangle\langle 0|\mathbf{V}^\pm_\text{eff}(E_{\alpha\pm1})|0\rangle\langle 0|}
	|0\rangle\langle 0|
\nonumber\\=&
	V^\pm_\text{eff}(E_{\alpha})|0\rangle\langle 0|
	.
\end{align}
%
In general, we compute these effective potentials numerically by setting a maximum frequency $E_M$ above which the effective potentials are zero, that is, $\mathbf{V}^\pm_\text{eff}(E_{m}) = 0 $, for all $E_m>E_M$. Typically we take $m$ to be on the order of 100, which normally is sufficient for convergence.

\section{Topological insulator surface}
First, we will investigate the effect of time-dependent impurity on the edge of three-dimensional topological insulators. The low energy effective model within $k\cdot p$ approximation  can be described  by
 \begin{align}
  H= v\sum_k
\Psi_\mathbf{k}^\dag[\mathbf{k}\times\vec{e}_3]\cdot\sigma\Psi_\mathbf{k}
 \end{align}
We are interested in the local properties of this system such as density of states $N({\bf r},\omega)$, which is related to the (retarded) Green function through the identity $N({\bf r},\omega)=-\tr\im{\bf G}^r({\bf r},{\bf r};\omega)/\pi$. As a function of the effective potential, we write the Green function in momentum space as
\begin{align}
G_{k,k'}=&
	\delta_{kk'}G_{k}^0
	+
	G^0_k
	\frac{V_\text{eff} }{1-V_\text{eff} G^0_0}
	G^0_{k'}
	,
\end{align}
where $G$, $G^0$, and $V$ are implicit functions of the energy.
 
\begin{figure}[t]
\centering
\includegraphics[width=0.8\columnwidth]{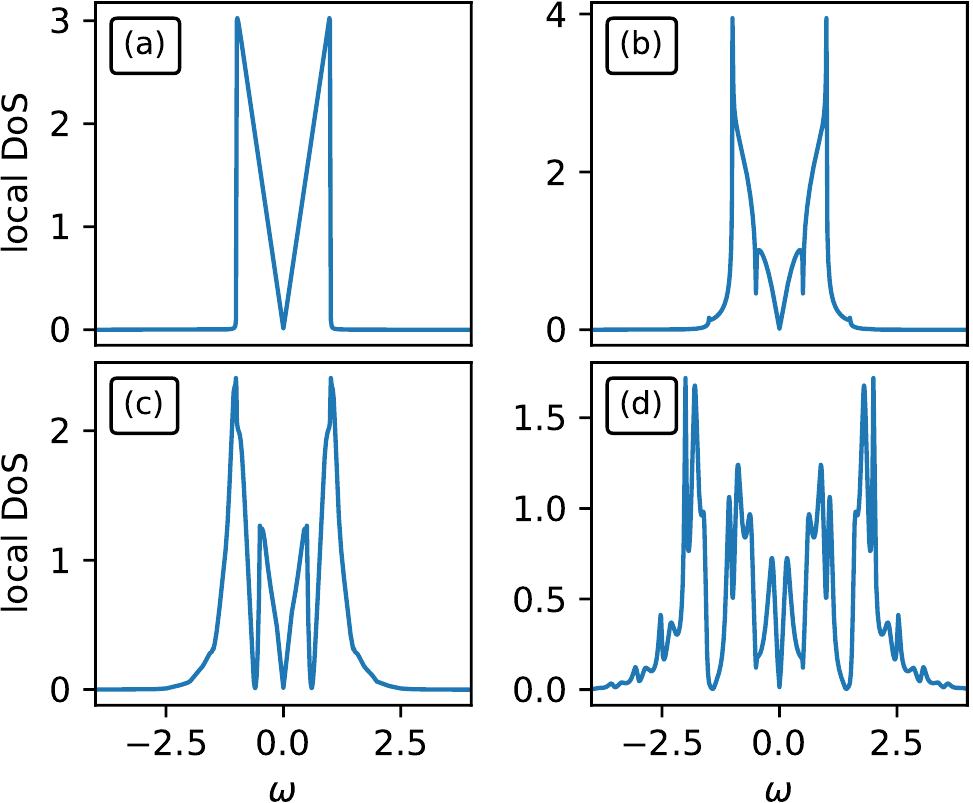}
 \caption{Local density of electron states at the impurity site with driving frequency $\Omega=0.50$  for different values of the driving amplitude $A =$ 0.0 (a), 0.5 (b), 1.0 (c), 2.0 (d). }
\label{fig:cont1}
\end{figure}

We consider the driving potential $ \mathbf{V}=2A\delta(\mathbf{r}-\mathbf{r}_0)\sigma^0\cos\Omega t$ with frequency $\Omega $, amplitude $A$, at the position $\mathbf{r}_0$. In Fig. \ref{fig:cont1}, we plot the local density of electron states for increasing impurity amplitude $A$, panels (a) through (d). In absence of the impurity, Fig. \ref{fig:cont1} (a), we retain the typical Dirac-like density of states with vanishing density of state at the Fermi energy, as expected. However, the linear density of states around the Fermi energy is preserved also at finite driving amplitudes, Fig. \ref{fig:cont1} (b) -- (d). In addition to the linear low energy spectrum, high energy features emerge with increasing $A$. These features, which appear symmetrically around the Dirac point, are direct consequences of the excitations that are generated by the time-dependent potential, much in analogy with the impurity side resonances that are caused by vibrational defects on Dirac materials \cite{PhysRevLett.110.026802,PhysRevB.87.245404}.
In fact, the symmetric appearance of the excitations is caused by the combination of positive and negative scattering potentials $\bfV^+_\text{efft}$ and $\bfV^-_\text{efft}$, respectively, each of which is responsible for the equally distributed set of excitations on either the valence \emph{or} the conduction side of the electronic band structure.
Moreover, the number of excitations increases substantially with increasing $A$, which we understand to be an effect the increased order of $\mathbf{C}_\alpha$ that contributes to the effective potential $\mathbf{V}_\text{eff}$. Therefore, the Green function picks up an increasing number of higher energy modes. We also notice that the bandwidth of the local density of states increases the stronger the driving force. We refer this to the increasing number of contributing excitations that become available for in the scattering processes and which necessitates a redistribution of the total density onto an increased set of excitations.

\section{Kane-Mele model}

In the second part of this article, we  will consider a lattice, Kane-Mele, model for a topological insulator. The model pertains to a tight-binding Hamiltonian  on a honeycomb lattice which is a straight forward generalization of the Haldane model \cite{haldane}. Here, we write
\begin{align}
H=&
	-t\sum_{\langle ij\rangle\sigma}c^\dag_{i\sigma}c_{j\sigma}
	+
	i\lambda_{so}\sum_{\langle\langle ij\rangle\rangle\sigma}c^\dag_{i\sigma}\sigma^zc_{j\sigma}
	+
	\mu \sum_{i\sigma}c^\dag_{i\sigma}c_{i\sigma}
	,
\end{align}
where $t$ is the nearest-neighbor (NN) hopping integral, $\lambda $ is the spin-orbit coupling which act as a next-nearest-neighbor hopping element, whereas $\mu $ is the chemical potential which fixes the number of particle of the system. Here, we consider a half-filled system for which $\mu=0$. Note that spin-orbit coupling act as a purely imaginary hopping integral and differ by a sign for up and down spin component. We set $t$ as our absolute energy scale, and we have imposed open boundary conditions in the $x$-direction and periodic boundary conditions along the $y$-direction. We calculate $V_\text{eff}$ iteratively for a lattice size $100\times50$. We consider the same driving potential as above, $\mathbf{V}=\delta(\mathbf{r}-\mathbf{r}_0)2A\sigma^0\cos\Omega t$, with the frequency $\Omega$,  amplitude $A$ at $\mathbf{r}_0$.

\begin{figure}[t]
\centering
\includegraphics[width=0.8\columnwidth]{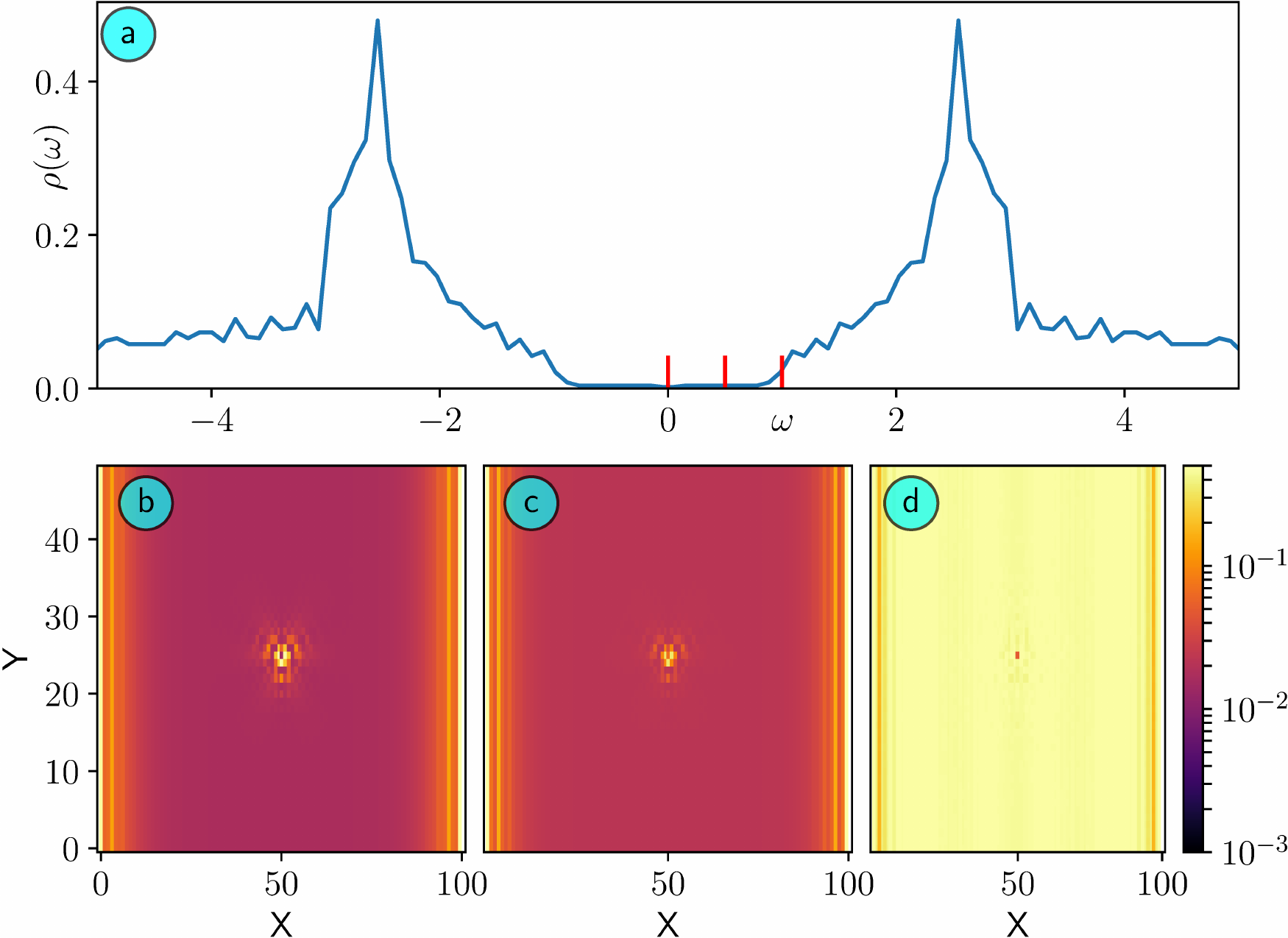}
\caption{(a) Total density of state of the Kane-Mele model for the  topological insulator with a single impurity at the middle of the lattice. The vertical lines indicate the energy at which local density of state is plotted in the bottom panel. Real space map of the local density of state for different  values of energy  E=0.0 (b), 0.5 (c), 1.0 (d).}
\label{fig:kane1}
\end{figure}

The plots in Fig. \ref{fig:kane1} display the total density of states of the
Kane-Mele model with a single impurity located at the center of the lattice. The
red vertical line signifies the energy values at which the spacial maps of the local density of state are plotted in panels (b) through (d). The density of states for this tight binding model on a honeycomb lattice vanishes at the Fermi energy and increases linearly away from the Fermi energy. A finite spin-orbit coupling opens up a gap in the spectrum. Due to the non-trivial topological nature of this system, this gap only appears in the bulk while metallic edge states appear at the boundary. In Fig. \ref{fig:kane1} (b), (c), this can be seen as an enhanced density of states at the edges $x=0$ and $x=100$. Moreover, one should notice the finite density of states around the impurity, a density which oscillates and decays far away from it. At higher energies, however, bulk states appear which displays a uniform density of states throughout the whole system, see Fig. \ref{fig:kane1} (d). 

\begin{figure}[b]
\centering
\includegraphics[width=0.8\columnwidth]{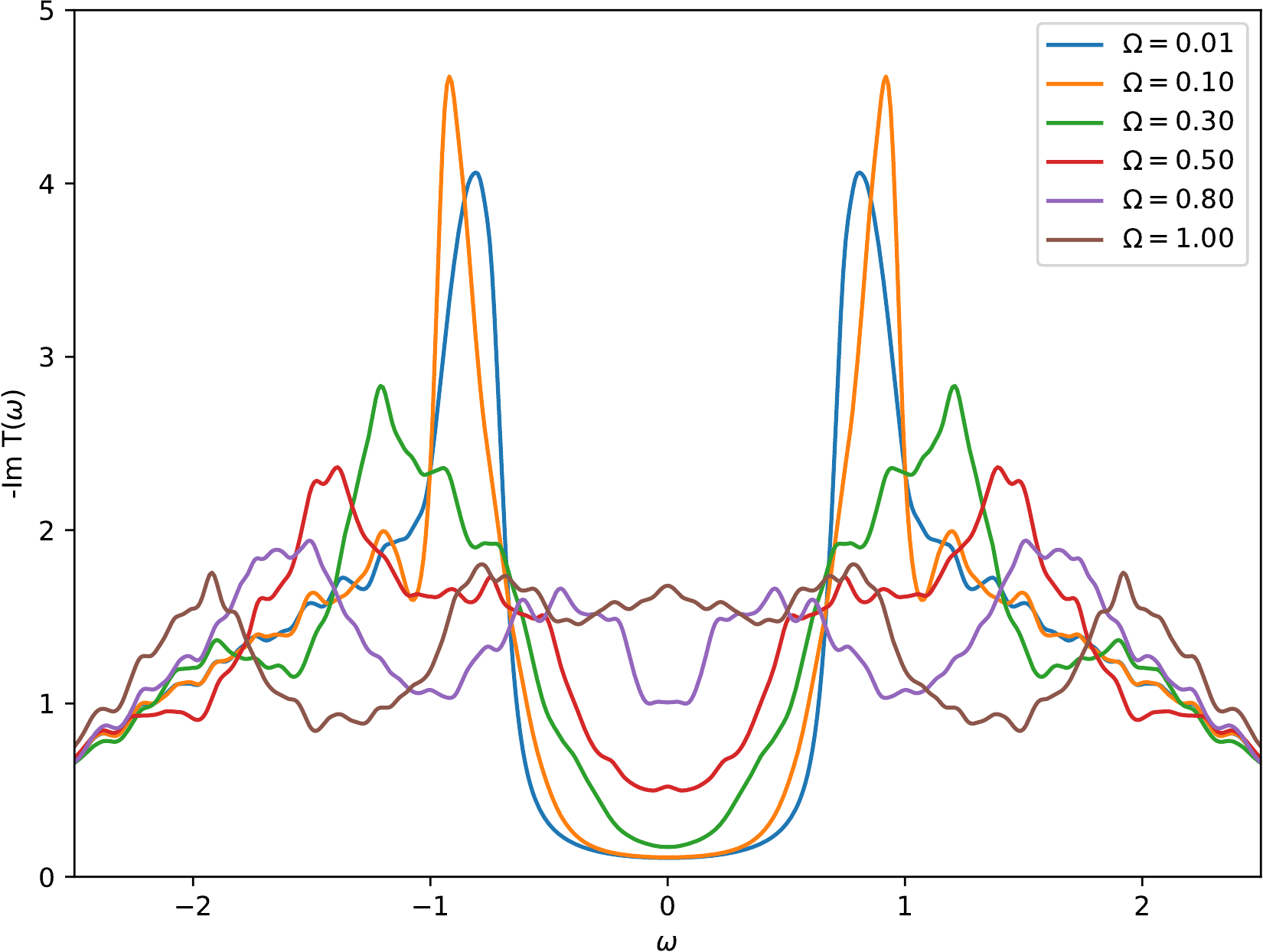}
\caption{$T$-matrix of an impurity at the middle of the lattice with driving
	amplitude $A = 5.00$. The bulk gap in the spectrum induced by finite spin-orbit coupling reflects in the $T$-matrix. When the driving frequency is small the gap remains intact and no states appear within the gap. With increasing driving frequency states appear in the gap.}
\label{fig:tmat1}
\end{figure}

The properties of the spectrum can be better understood by considering the imaginary part of the $T$-matrix, defined by
\begin{align}
{\cal T}=&
	\frac{V_\text{eff} }{1-V_\text{eff} G^0_0}
	,
\end{align}
as a function of energy. In Fig. \ref{fig:tmat1} we plot $-\im\, {\cal T}$ as  a function of the energy for six different frequencies $\Omega$ of the driving field. The $T$-matrix contains scattering effects of all orders, and the plots in Fig. \ref{fig:tmat1} show $-\im\, {\cal T}$ for an impurity located at the center of the lattice. Since the bulk of a topological insulator does not have any states available near the Fermi energy, the imaginary part of the $T$-matrix remains gapped for small driving frequencies. Hence, the spectral content of the $T$-matrix only marginally deviates from the expected spectral properties of the pristine lattice. However, huge resonances appear symmetrically outside the bulk gap, providing coherent resonance peaks. These resonances can be thought of as in similar terms as the impurity resonances induced in Dirac materials by particle scattering off  local defects \cite{AdvPhys.63.1,Nature.403.746,PhysRevLett.104.096804,PhysRevB.81.233405,PhysRevB.85.121103,edge6,PhysRevB.94.075401}.

By contrasts, it can be noticed from the plots in Fig. \ref{fig:tmat1} that, with increasing frequency of the driving field, a finite number of states emerge within the bulk gap which tends to become filled up by these states. Simultaneously, the coherent peaks vanish with increased driving frequency as a result of the necessary charge redistribution which follows from the emergence of additional resonances in the spectrum.

\begin{figure}[t]
\centering
\includegraphics[width=0.8\columnwidth]{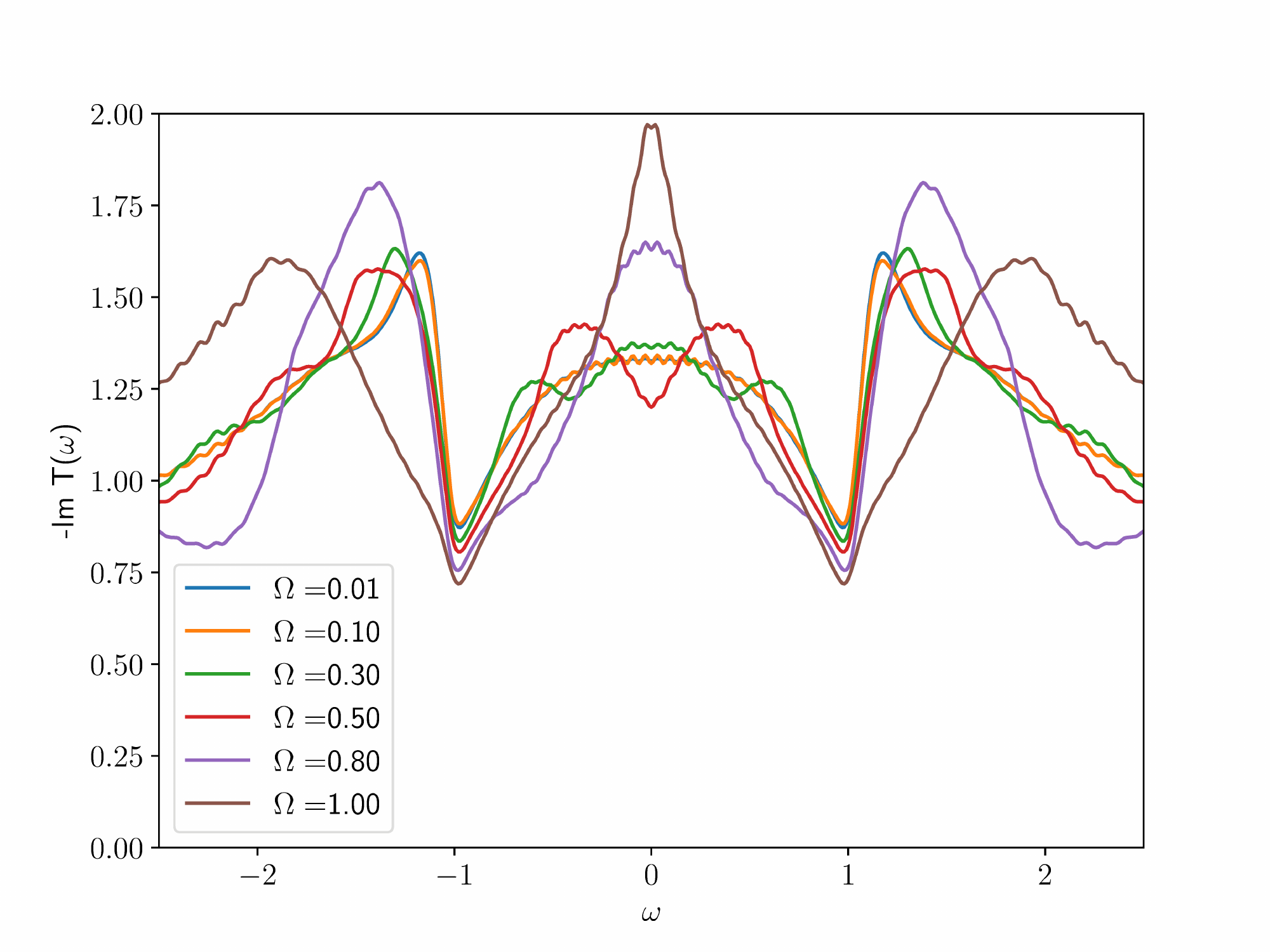}
\caption{$T$-matrix of an impurity at the edge of the lattice with driving amplitude  $A = 0.50$.  Topological Insulators have edge state at the boundary of the sample and a finite no of scattering states appear within the bulk gap. $T$-matrix also, show similar behavior with a finite weight within the gap even for small frequency region.}
\label{fig:tmat2}
\end{figure}

Next, we consider the impurity to be located at the edge of the lattice, that
is, $(x,y)=(0,25)$, resulting in the plots $-\im\, {\cal T}$ shown in Fig.
\ref{fig:tmat2} for six different frequencies of the driving field. In stark contrast to the situation discussed, the coherence peaks outside the bulk gap become are turned into dips. The metallic nature of the edge states gives rise to finite scattering states even for small driving frequencies. As the driving frequency is increased, an increased number of Floquet modes contribute to the scattering process, which leads to a non-monotonic oscillation pattern in the $T$-matrix.  

\section{Local density of states}

\begin{figure}[t]
\centering
\includegraphics[width=0.8\columnwidth]{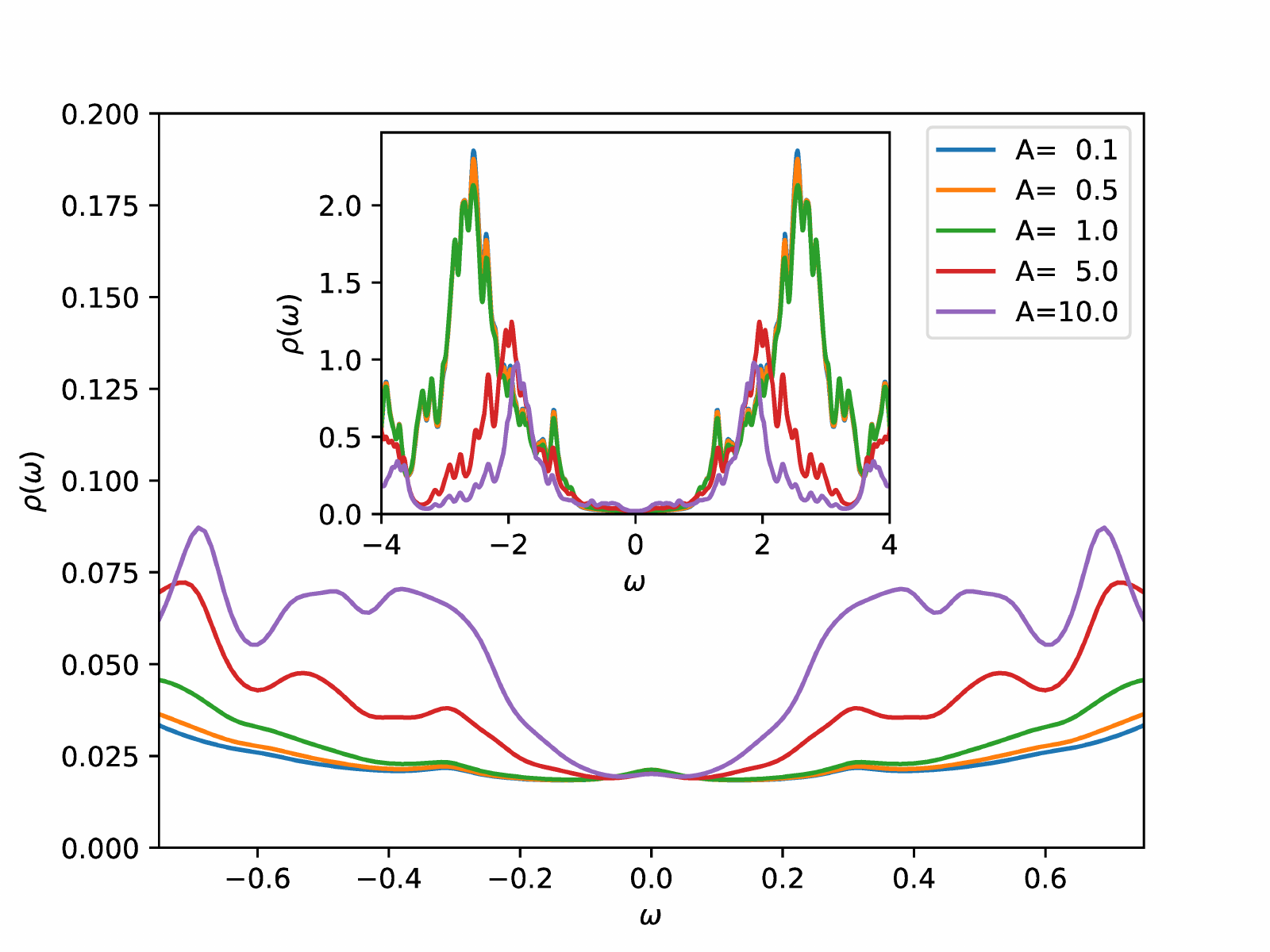}
\caption{Local density of electron states for a single impurity with driving
	frequency $\Omega=0.50$ at the impurity site which is located in the
	middle of the lattice for small energy. The inset figure shows the local density of electron states in a large energy range.}
\label{fig:ldos1}
\end{figure}

In this section, we consider the effect of impurity on the local density of electron 
states, something which might be useful for local probing experiments, such as scanning 
tunneling microscopy. First, we focus on the case where the impurity is located
at the  
center of the lattice. In Fig. \ref{fig:ldos1}, we plot the local density of electron states at the impurity site for a sequence of 
different values of the amplitude $A$ of the driving field for a fixed frequency $\Omega$.  
In the absence of the impurity, $A=0$, the local density of states at the center (in bulk, far from the edge) of the lattice 
has a gap (not shown), which is caused by the finite spin-orbit coupling.
For a weak driving amplitude, the local density of electron states shows 
a hard gap, which is reminiscent of the bulk gap in the unperturbed case.
Surprisingly, however, even by increasing the driving amplitude by more than an order 
of magnitude, the local density of electron states does not change appreciably within the 
gap. This difference is in stark contrast when compared to the T-matrix results, where a finite weight 
appears when the impurity is located at the center of the lattice. 
The local density of electron states increases away from the Fermi energy, which can be seen by zooming in inside the gap, while the local density of electron states remains constant at the Fermi energy.  This implies that even though 
there is a finite number of states appearing within the bulk gap due to impurity scattering, 
the density of states at the Fermi level does not change. The coherence peaks emerging at the edges of the gap, become less prominent with increasing amplitude of the driving field.

\begin{figure}[t]
\centering
\includegraphics[width=0.8\columnwidth]{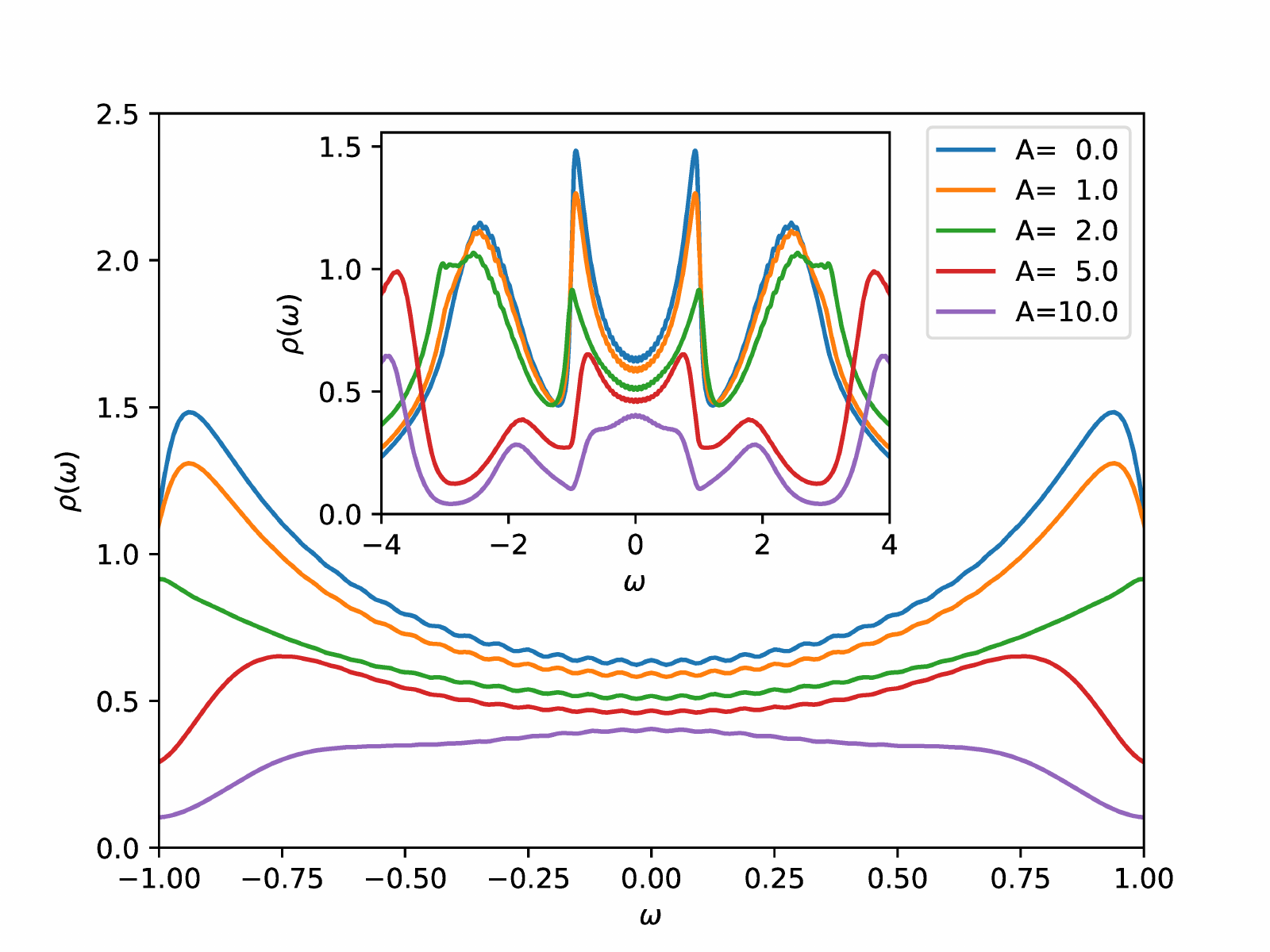}
\caption{Local density of electrons states for a single impurity with driving frequency $\Omega = 2.00$ at the impurity site which is located at the edge of the lattice for small energy. Inset figure show local density of electron states in a large energy range.}
\label{fig:ldos2}
\end{figure}

In Fig. \ref{fig:ldos2}, we plot the local density of electron states for the case where the impurity is located at 
the edge of the lattice. It is clear that this set-up leads to completely different kinds of features in 
the local density of electron states, as compared to the previous one.
The non-trivial nature of the metallic TI metallic 
edge states, due to the chirality, becomes more significant at the boundary of
the lattice. As can be seen in Fig. \ref{fig:ldos2}, the local density of electron states is not gapped in the absence of impurities, $A=0$. This property is preserved for weak driving fields, $0<A\lesssim2$, although it is also clear that both the density of states at the Fermi level as well as the sharp resonances ($|\omega|\approx1$) become slightly weaker, there is, nonetheless, a finite local density of electron states inside the bulk gap. With even larger amplitudes, the overall low energy local density of electron states becomes simultaneously broadened and suppressed in amplitude.

\section{Summary and conclusions} 
In this paper, we have studied the effects of a time-dependent potential 
impurity in two-dimensional topological insulators. By employing a numerically 
exact  matrix continued fraction method, which takes into account higher order 
Floquet modes, we capture the low energy physics throughout a wide range between 
weak and strong driving fields. This approach is implemented for both continuum 
and lattice models of the topological insulator.

Regarding the continuum model, we find that the low lying energy states remain unaffected by driving fields with weak to moderate strength, whereas the higher lying energy states become strongly modified. The linear property of the local density of electron states disappears with increasing amplitude of the driving field. Many more oscillations emerge in the local density of electron states due to contributions from the higher Floquet modes at larger driving amplitude, as well as a slight increase in the bandwidth.

We have imposed open boundary condition for the two-dimensional lattice model of
the topological insulator along with $x$-direction, which gives metallic edge
states. Here, we have analyzed the effect of an impurity at two different
locations, namely, at the center and an edge of the lattice. The bare local
density of electron states has insulating and metallic properties at these two lattice sites, respectively. The imaginary part of the $T$-matrix has very different responses for impurities located at these two positions. For a weakly perturbing impurity located at the center of the lattice, the bulk gap of the lattice is preserved in the $T$-matrix, while a finite density grows up within the gap at stronger driving fields caused by increased impurity scattering. For strong enough fields, this density eventually fills up the gap.
On the other hand for an impurity located at the edge of the lattice, the $T$-matrix and, hence, the impurity scattering, largely impacts the electronic density around the Fermi energy already for weak driving fields.

We conclude that despite the apparent simplicity of the model and some
similarities with the physics originating from single impurity interacting with
the surfaces states of a three dimensional topological insulator, there is
increased complexity in the present set-up which can be directly linked to the
time-dependent potential. This should have bearing on issues related to vibrational defects as well as the external driving field we have in mind here. However, since the driving frequencies in the two situations should be expected to differ by orders of magnitude, we can, nonetheless, conjecture that our low frequency results should be applicable to vibrational impurities. We anticipate that studies of the time-dependent using different methods should corroborate our findings.

\acknowledgments
We thank Stiftelsen Olle Engqvist Byggm\"astare and Vetenskapsr\aa det for financial support.


\begin{thebibliography}{200}
\bibitem{chaos1} D.F. Escande,  Phys. Rep \textbf{121}, 165 (1985) 
\bibitem{chaos2} L.E. Reichl and  W.M. Zheng,  Directions in chaos, \textbf{1}, Hao Bailin (ed.). Singapore: World Scientific 1987 
\bibitem{chaos3} A.J. Lichtenberg and  M.A. Lieberman, Regular and stochastic motion. New York: Springer 1981 
\bibitem{chaos4} G.M. Zaslavsky,  Phys. Rep. \textbf{80},157 (1981) 
\bibitem{Fainshtein} A. G. Fainshtein, N. L. Manakov, V. D. Ovsiannikov, and L. P. Rapoport, Phys. Rep. 210, 111 (1992).
\bibitem{coldatom1} F. L. Moore, J. C. Robinson, C. Bharucha, P. E. Williams, 
and M. G. Raizen, Phys. Rev. Lett. 73, 2974 (1994);
\bibitem{coldatom2} P. J. Bardroff, I. Bialynicki-Birula, D. S. 
Kr\"{a}hmer, G. Kurizki, E. Mayr, P. Stifter, and W. P. Schleich, 
Phys. Rev. Lett. \textbf{74} 3959 (1995).

\bibitem{drive1} N. H. Lindner, G. Refael, and V. Galitski, Nat. Phys. \textbf{7}, 490(2011).
 \bibitem{drive2} T. Oka and H. Aoki, Phys. Rev. B \textbf{79}, 081406 (2009).
 \bibitem{drive3} T. Kitagawa, T. Oka, A. Brataas, L. Fu, and E. Demler, Phys. Rev. B \textbf{84}, 235108 (2011).
 \bibitem{drive4} L. Jiang, T. Kitagawa, J. Alicea, A. R. Akhmerov, D. Pekker, G. Refael, J. I. Cirac, E. Demler, M. D. Lukin, and P. Zoller, Phys. Rev. Lett. \textbf{106}, 220402 (2011).
 \bibitem{drive5} N. H. Lindner, D. L. Bergman, G. Refael, and V. Galitski, Phys. Rev. B \textbf{87}, 235131 (2013).
 \bibitem{drive6} Y. T. Katan and D. Podolsky, Phys. Rev. Lett. \textbf{110}, 016802 (2013).
 \bibitem{drive7} G. Usaj, P. M. Perez-Piskunow, L. E. F. Foa Torres, and C. A. Balseiro, Phys. Rev. B \textbf{90}, 115423 (2014).
 \bibitem{drive8} Z. Gu, H. A. Fertig, D. P. Arovas, and A. Auerbach, Phys. Rev. Lett. \textbf{107}, 216601 (2011).
 \bibitem{drive9} D. E. Liu, A. Levchenko, and H. U. Baranger, Phys. Rev. Lett. \textbf{111}, 047002 (2013).
 \bibitem{drive10} Y. Li, A. Kundu, F. Zhong, and B. Seradjeh, Phys. Rev. B \textbf{90}, 121401 (2014).

\bibitem{haldane} C. L. Kane and E. J. Mele, Phys. Rev. Lett. \textbf{95}, 146802 (2005).
(2005).

 
 \bibitem{shirley} J. H. Shirley, Phys. Rev. \textbf{138}, B979 A965).

 \bibitem{ftimat1} Y. H. Wang, H. Steinberg, P. Jarillo-Herrero, and N. Gedik, Science \textbf{342}, 453 (2013).
 \bibitem{ftimat2} F. Mahmood, C.-K. Chan, Z. Alpichshev, D. Gardner, Y. Lee, P. A. Lee, and N. Gedik, Nat. Phys. \textbf{12}, 306 (2016).
 \bibitem{ftiopt} M. C. Rechtsman, J. M. Zeuner, Y. Plotnik, Y. Lumer, D. Podolsky, F. Dreisow, S. Nolte, M. Segev, and A. Szameit, Nature (London) \textbf{496}, 196 (2013).
 \bibitem{fticold} G. Jotzu, M. Messer, R. Desbuquois, M. Lebrat, T. Uehlinger, D. Greif, and T. Esslinger, Nature (London) \textbf{515}, 237 (2014).
\bibitem{ti1} H. Zhang, C.-X. Liu, X.-L. Qi, X. Dai, Z. Fang, and S.-C. Zhang,
Nat. Phys. \textbf{5}, 438 (2009).
\bibitem{ti2} Y. L. Chen \textit{et al.}, Science \textbf{325}, 178 (2009).
\bibitem{ti3} Y. Xia \textit{et al.}, Nat. Phys. \textbf{5}, 398 (2009).
(2007).
\bibitem{edge1}  W.-C. Lee, C. Wu, D. P. Arovas, and S.-C. Zhang, Phys. Rev. B 
\textbf{80},245439 (2009); 
\bibitem{edge2} X. Zhou, C. Fang, W.-F. Tsai, and J. P. Hu, Phys. Rev. B  
\textbf{80}, 245317 (2009); 
\bibitem{edge3} H.-M. Guo and M. Franz,  Phys. Rev. B \textbf{81}, 041102 
(2010);
\bibitem{edge4} A. M. Black-Schaffer and A. V. Balatsky, Phys. Rev. B 
\textbf{85}, 121103(R) (2012);
\bibitem{edge5} J. Fransson, A. M. Black-Schaffer, and A. V. Balatsky, Phys. 
Rev. B, \textbf{90}, 241409(R) (2014);
\bibitem{edge6}  A. M. Black-Schaffer, A. V. Balatsky, and J. Fransson, Phys. 
Rev. B \textbf{91}, 201411(R) (2015).
\bibitem{PhysRevB.94.075401} J. Fransson, A. M. Black-Schaffer, and A. V. Balatsky  Phys. Rev. B {\bf 94}, 075401 (2016).

\bibitem{edge7}  P. Roushan, J. Seo, C. V. Parker, Y. S. Hor, D. Hsieh, D. Qian,
A. Richardella, M. Z. Hasan, R. J. Cava, and A. Yazdani, Nature
(London) 460 (2009); 
 \bibitem{edge8} T. Zhang, P. Cheng, X. Chen, J.-F. Jia,
X. Ma, K. He, L. Wang, H. Zhang, X. Dai, Z. Fang et al., Phys. Rev.
Lett. \textbf{103}, 266803 (2009); 
\bibitem{edge9} Z. Alpichshev, J. G. Analytis, J.-H. Chu,
I. R. Fisher, Y. L. Chen, Z. X. Shen, A. Fang, and A. Kapitulnik,
Phys. Rev. Lett.  \textbf{104}, 016401 (2010).

\bibitem{mag-im1} A.  Roth,  C.  Br\"{u}ne,  H.  Buhmann,  L.  W.  Molenkamp,
J. Maciejko, X.-L. Qi, and S.-C. Zhang, Science \textbf{325} , 294 (2009);
\bibitem{mag-im2} M.  K\"{o}nig,  S.  Wiedmann,  C.  Br\"{u}e,  A.  Roth,  H.  
Buhmann,  L.  W.  Molenkamp,  X.-L.  Qi,  and  S.-C.  Zhang, Science 
\textbf{318}, 766 (2007);
\bibitem{mag-im3} G. M. Gusev, Z. D. Kvon, O. A. Shegai, N. N. Mikhailov,
S.  A.  Dvoretsky,  and  J.  C.  Portal,  Phys.  Rev.  B \textbf{84}, 121302(R) 
(2011).

\bibitem{lesser1} G. Casati, I. Guarneri, D. L. Shepelyansky, IEEE J. Quant. 
Elect. 24, 1420 (1988);
\bibitem{lesser2} R. Shakeshaft, Comm. At. Mol. Phys. \textbf{28}, 179 (1992);
\bibitem{lesser3} Laser Phys. 3, No. 2 (1993), special issue on Atoms, Ions 
and Molecules in a Strong Laser Field, edited by A. M. Prokhorov.
\bibitem{Floquet1} G. Floquet, Ann. de l’Ecole Norm. Sup. \textbf{12}, 47 
(1883);
\bibitem{Floquet2} E. L. Ince, Ordinary Differential Equations (Dover Publ., 
New York, 1956);
\bibitem{Floquet3} W. Magnus and S. Winkler, Hill’s Equation (Dover Publ., New 
York, 1979).


 \bibitem{bernevig} C. Wu, B. A. Bernevig, and S.-C. Zhang, Phys. Rev. Lett. \textbf{96},106401(2006).
 \bibitem{moore}C. Xu and J. E. Moore, Phys. Rev. B \textbf{73}, 045322 (2006).



\bibitem{Risken}Risken. H, The Fokker-Planck equation. Springer Series in Synergetics, Vol. 18. Berlin, Heidelberg, New 
York: Springer 1984 
\bibitem{martinez1} D F Martinez,  J. Phys. A: Math. Gen. \textbf{36} 9827 
(2003). 
\bibitem{martinez2} F. Grossmann, P. Jung, T. Dittrich, and P. H\"{a}nggi, Z. 
Phys. B \textbf{84}, 315 (1991).

\bibitem{PhysRevLett.110.026802} J. -H. She, J. Fransson, A. R. Bishop, and A. V. Balatsky, Phys. Rev. Lett. {\bf 110}, 026802 (2013).
\bibitem{PhysRevB.87.245404} J. Fransson, J -H. She, L. Pietronero, and A. V. Balatsky, Phys. Rev. B, {\bf 87}, 245404 (2013).

\bibitem{AdvPhys.63.1} T. Wehling, A. Black-Schaffer, and A. Balatsky, Adv. Phys. {\bf 63}, 1 (2014).
\bibitem{Nature.403.746} S. H. Pan, E. W. Hudson, K. M. Lang, H. Eisaki, S. Uchida, and J. C. Davis, Nature (London) {\bf 403}, 746 (2000).
\bibitem{PhysRevLett.104.096804} M. M. Ugeda, I. Brihuega, F. Guinea, and J. M. G\'omez-Rodr\'iguez, Phys. Rev. Lett. {\bf 104}, 096804 (2010).
\bibitem{PhysRevB.81.233405} R. R. Biswas and A. V. Balatsky, Phys. Rev. B {\bf 81}, 233405 (2010).
\bibitem{PhysRevB.85.121103} A. M. Black-Schaffer and A. V. Balatsky, Phys. Rev. B {\bf 85}, 121103 (2012).


\end{thebibliography}
\end{document}